\documentclass{iau} 
\usepackage{graphicx}
\usepackage{color}

          \font\sixrm=cmr6

{\catcode`\@=11                                                  
\gdef\SchlangeUnter#1#2{\lower2pt\vbox{\baselineskip 0pt\lineskip0pt    
\ialign{$\m@th#1\hfil##\hfil$\crcr#2\crcr\sim\crcr}}}}           
\def\gtrsim{\mathrel{\mathpalette\SchlangeUnter>}}               
\def\lesssim{\mathrel{\mathpalette\SchlangeUnter<}}    
\def\teq#1{$\, #1\,$}                         
\def\dover#1#2{\hbox{${{\displaystyle#1 \vphantom{(} }\over{
   \displaystyle #2 \vphantom{(} }}$}}
\def\Machalf{{\cal M}_{\hbox{\sixrm A}}}

\title[IAUS 331.~~X-ray Synchrotron Polarization in SNRs] 
{X-ray Synchrotron Polarization from Turbulent Plasmas in Supernova Remnants}

\author[Matthew G. Baring]   
{Matthew G. Baring}

\affiliation{Department of Physics and Astronomy - MS 108, Rice University, \\
6100 Main Street, Houston, Texas 77251-1892, USA
\\email: {\tt baring@rice.edu}}

\pubyear{2017}
\volume{331}  
\jname{SN 1987A, 30 years later --- Cosmic Rays and Nuclei from \\Supernovae and their aftermaths}
\editors{M. Renaud, A. Marcowith, G. Dubner, A. Ray \& A. Bykov, eds.}
\begin{document}

\maketitle

\begin{abstract}
As supernova remnants (SNRs) age, they become
efficient cosmic ray accelerators at their outer shell shocks.  The
current paradigm for shock acceleration theory favors turbulent field
environs in the proximity of these shocks, turbulence driven by current
instabilities involving energetic ions.  With the
imminent prospect of dedicated X-ray polarimeters becoming a reality,
the possibility looms of probing turbulence on scales that couple to the
super-TeV electrons that emit X-rays.  This paper presents model X-ray
polarization signatures from energetic electrons moving in simulated MHD
turbulence of varying levels of ``chaos.''  The emission volumes are
finite slabs that represent the active regions of young SNR
shells.  We find that the turbulent field energy must be
quite limited relative to that of the total field in order for the X-ray
polarization degree to be as strong as the radio measures obtained in
some remnants.  Results presented are
pertinent to the planned IXPE and XIPE polarimeters.
\end{abstract}

\firstsection 

\section{Introduction}

One of the most important contributions of the {\it Chandra} X-ray Observatory 
to supernova remnant (SNR) science has been the discovery of thin synchrotron rims
in the outer shells of several remnants.  This advance was enabled by 
{\it Chandra}'s exquisite imaging capability. 
Among these sources are SN1006 (Long et al. 2003) and Cassiopeia A 
(Vink \& Laming 2003), the original {\it first light} target for the Observatory.
The sharp rise and fall of the X-ray flux in these filamentary rims, on 
angular scales of a few arcseconds, implies a very short synchrotron cooling 
timescale.  This yields the unavoidable interpretation that the embedded 
magnetic field is on the order of 20--70$\mu$Gauss, substantially above 
values normally interpreted from standard invocations of field compression 
in MHD shocks.  Thus a quandary emerged: how could such anomalous 
{\bf field enhancements} come about?

In the years just prior to and around this time, the concept of {\bf turbulent amplification} 
of magnetic fields mediated by cosmic ray-driven instabilities in SNR shocks 
was being developed, led by the papers by Lucek \& Bell (2000) and Bell (2005).
The anisotropy of the most energetic ions on the scales of the thickness of remnant 
shells seeds growth of MHD fluctuations to the levels of \teq{(\delta B/B)^2
\sim \Machalf P_{\rm cr}/\rho u^2 \gtrsim 3-10}.  Here 
\teq{\Machalf} is the Alfv\'enic Mach number, \teq{P_{\rm cr}} is the 
cosmic ray (CR) pressure, and \teq{\rho u^2} is the fluid ram pressure.  
Basically, the CR pressure gradient does work on the field energy density.
The interplay between the thermal gas, the cosmic rays and the field energy 
introduces inherently non-linear elements to the dynamics of the shock environs
(e.g. Vladimirov, Ellison \& Bykov 2006).
The observation of thin rims propelled the extensive development of models of the 
so-called {\bf Bell instability}, and the paradigm that SNR shells 
contain fields that are turbulent and enhanced has become very popular.

\newpage

We are now at the dawn of a new era in X-ray imaging, namely with the 
advent of X-ray polarimetry.  NASA announced earlier this year the 
selection of the Imaging X-ray Polarimetry Explorer (IXPE, see 
Weisskopf et al. 2013) as a Small Explorer mission, to be launched in four or 
so years.  This will serve as a pathfinder for discovery of polarized sources in 
X-rays, with a special focus on imaging and therefore on supernova remnants.
IXPE's nominal \teq{30''} angular resolution will be able to probe the diffusive scales 
of the most energetic electrons emitting synchrotron X-rays in SNRs. It will 
thus be able to cast light on the level and nature of turbulence in SNR rims 
and shells, with around 8 prime SNR targets in its advertised science program
(Weisskopf et al. 2013).\footnote{See also the IXPE web page at {\tt https://wwwastro.msfc.nasa.gov/ixpe/}.}
Subsequent X-ray polarimetry initiatives such as XIPE (Soffitta et al. 2013)
and eXTP offer additional prospects for probing the SNR environment in the 
not too distant future.

In preparation for such a data influx, exploration of what one might expect 
to see with an X-ray polarimeter is strongly motivated.  A seminal study 
of the flux and polarization degree variations in turbulent field models 
of synchrotron-emitting remnants was published in Bykov et al. (2009), 
outlining how images sampling various angular scales can capture the 
information of field turbulence.  Here we present results from an incipient 
simulation of synchrotron radiation from charges in prescribed 
MHD turbulence using a somewhat different construction, but notably 
with a decidedly different emphasis.  Here the simulation output 
provides measures of the fluctuations, mean and standard deviation of 
the polarization Stokes parameters, and demonstrates the strong 
correlation between these and the variance of Kolmogorov turbulence
in SNR shells.

\section{Synchrotron Polarization from Electrons in Magnetic Turbulence}

\begin{figure}[h]
\begin{center}
 \includegraphics[width=2.5in]{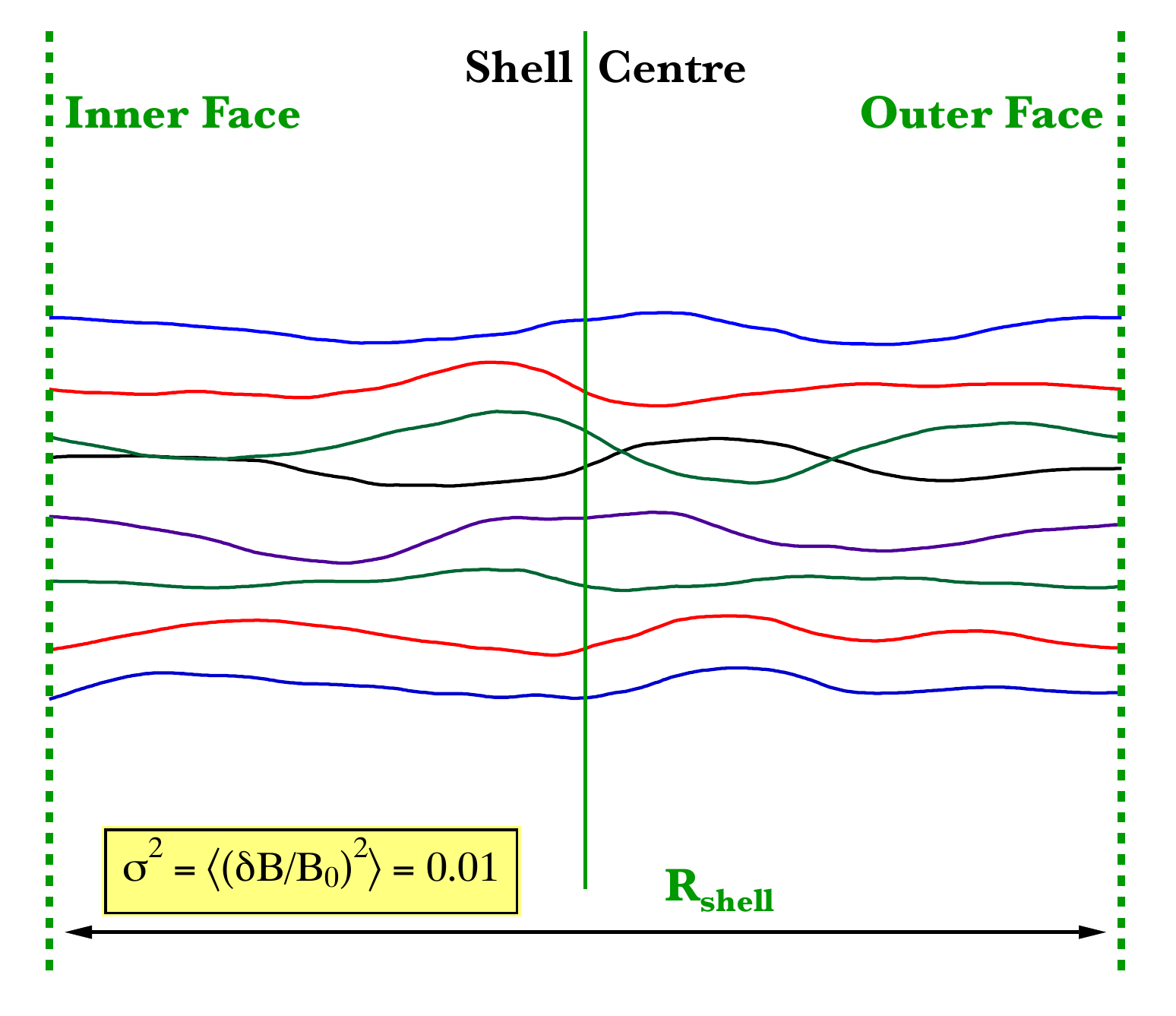} 
 \hspace{-3pt}
 \includegraphics[width=2.5in]{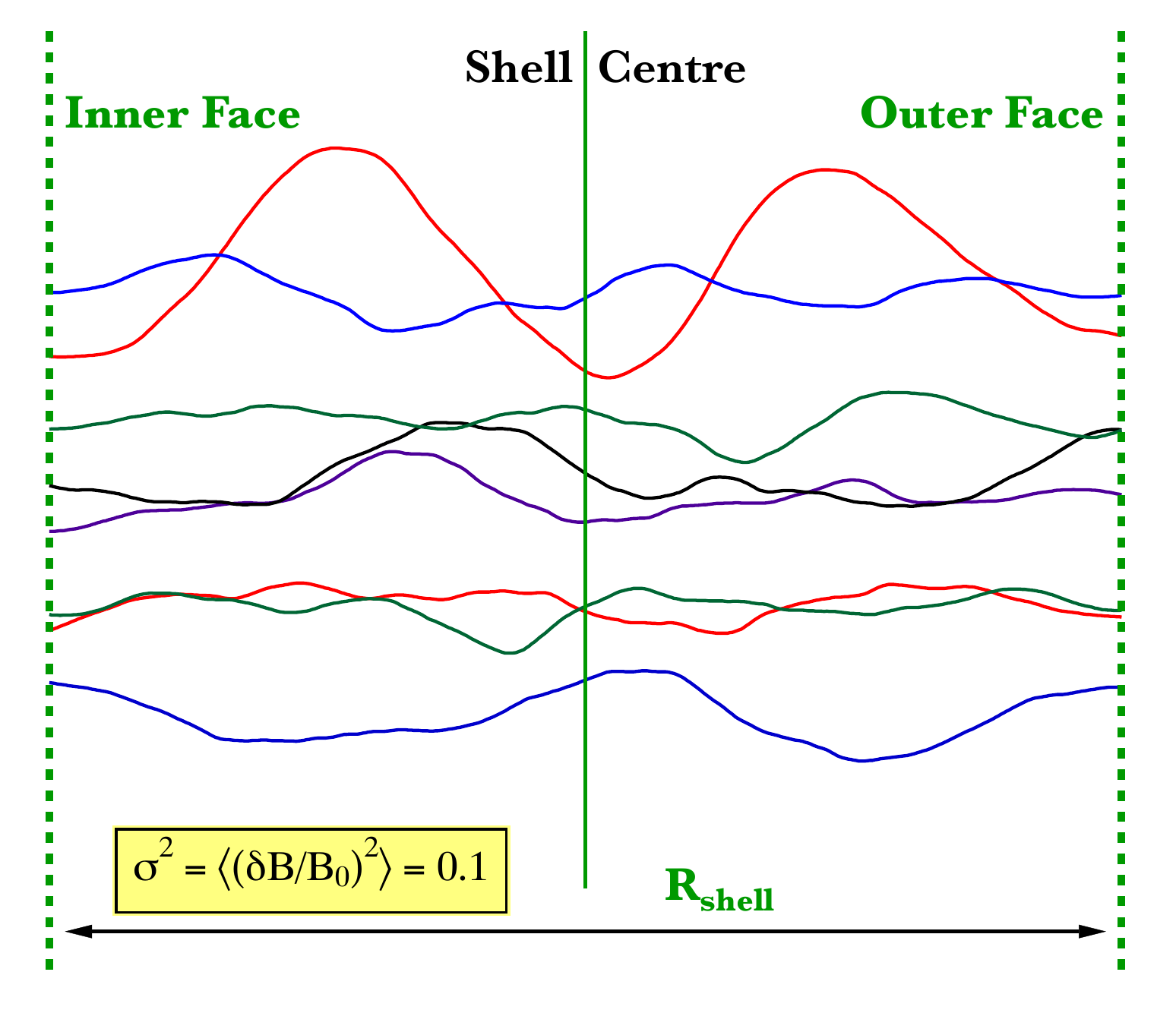} 
 \caption{Schematics for the geometry of SNR shell containing turbulent fields.  
 The depicted field lines are from the prescription of Kolmogorov slab turbulence
 in Eq.~(\ref{eq:power_spec}), with the variance \teq{\sigma^2 = \langle (\delta B/B_0)^2\rangle} of the  
 turbulence at the stirring scale \teq{R_{\rm shell}/3} being \teq{\sigma^2=10^{-2}}
 for the left panel, and \teq{\sigma^2=10^{-1}} for the right panel.  Only the 
 projections of the field lines in the plane of the Figure are represented; the color 
 coding serves only to aid distinguishability of the lines.}
   \label{fig1:snr_turb}
\end{center}
\end{figure}

To assess the nature of polarized synchrotron emission
in turbulence near SNR shocks, a simulation of charge motion in prescribed 
magnetostatic field fluctuations was developed.  The construction was 
similar to the approach of Giacalone \& Jokipii (1999) in their charge diffusion study, 
in that 1D turbulence was superposed upon a uniform background field 
\teq{\mathbf{B}_0}, which is radially-directed along the \teq{x}-axis.  
The MHD turbulence assumes the form
\begin{equation}
   \delta \mathbf{B} \; =\;\sum_k \delta B_k \, 
   \Bigl\{ \cos \phi_{\perp}\, \hat{y} + \sin \phi_{\perp}\, \hat{z} \Bigr\}
   \, \cos \bigl( k x + \phi_k \bigr) \quad .
 \label{eq:fluct_form}
\end{equation}
The slab nature is embodied in the property that the wavenumbers {\bf k} of 
turbulence only possess a component \teq{k \hat{x}} along the direction 
of the unperturbed field \teq{\mathbf{B}_0}, and the transverse field 
perturbations are in the \teq{(y,\, z)}-plane, perpendicular to the 
plane of the sky (adopted for simplicity though not necessarily true, 
it is a testable assumption nonetheless).  The wave phase \teq{\phi_k}
and the azimuthal orientation \teq{\phi_{\perp}} were chosen
randomly. The power spectrum of the fluctuations was presumed to have 
a 1D Kolmogorov form:
\begin{equation}
   \dover{\langle \delta B_k \rangle^2}{B_0^2} \; =\; \sigma^2\, \left( \dover{k}{k_-} \right)^{-5/3}
   \quad ,\quad
   k_- \;\leq\; k \;\leq\; k_+\quad .
 \label{eq:power_spec}
\end{equation}
Here \teq{\sigma^2} is the wave variance at the base of the inertial range,
i.e. at \teq{k=k_- \equiv 2\pi /\lambda_{\rm stir}}, with \teq{\lambda_{\rm stir}
\sim R_{\rm shell}/3} being the stirring scale for the cascading turbulence.
Also, for the simulations herein,  \teq{k_+=100k_-} was adopted with
values of \teq{k} sampled randomly on a logarithmic scale over the inertial range.
Representative simulated turbulent fields are illustrated in Fig.~\ref{fig1:snr_turb}
via their projections onto the sky plane, which is taken to encapsulate
the radial vector through the slab.
The absence of any time dependence (magnetostatic assumption) 
in \teq{\delta \mathbf{B}} is a tantamount to presuming non-relativistic 
plasma dynamics.

Yet the synchrotron-radiating electrons are indeed ultra-relativistic.  They were injected 
at the inner boundary (\teq{x=-10}) of the slab, distributed randomly 
in the \teq{(y,\, z)} plane, and with the initial momentum vector directions selected
randomly from an isotropic distribution.   The Lorentz factors were selected 
randomly on a logarithmic scale from a power law distribution
over a narrow range spanning a factor of two in \teq{\Gamma_e}.  
The actual scale was a convolution of \teq{\Gamma_e/\sqrt{B_0}} so as 
to position the synchrotron spectral information neatly in the X-ray band.
The electrons were then propagated into the slab, solving the Lorentz force 
equation in the turbulent field structure in incremental intervals along 
their trajectories using a fourth-order Euler ODE solver.  Their paths 
and acceleration vectors were logged for use in the polarization algorithms.

\subsection{Synchrotron Polarization Characteristics}

To develop a general idea of the properties of synchrotron radiation polarization 
in SNR shells, the synchrotron emissivity is integrated over the charges'
trajectories.  This polarized emissivity is principally dependent on three quantities 
at each point along the lepton path: the magnetic field vector and the electron's 
Lorentz factor and pitch angle.  Since the charges were injected into the 
slab with an isotropic angular distribution, provided the turbulence is not 
too large, they retain approximate isotropy throughout the slab; this is relinquished somewhat 
when \teq{\delta B/B \sim 1}.  The local acceleration vector determines the 
instantaneous pitch angle, and combined with {\bf B}, they serve as 
parametric input for the standard synchrotron emissivity formulae 
(e.g. see \cite[Rybicki \& Lightman 1979]{rl79}).  The electron Lorentz factors are distributed, and 
for our purposes here, a power-law distribution \teq{n_e(\gamma) \propto \gamma^{-p}}
with \teq{p =3} was presumed.  This generates a synchrotron intensity spectrum 
\teq{I_{\nu}\propto \nu^{-1}}, which approximates the steep synchrotron 
slopes in the keV X-ray band seen in SNRs that are interpreted as portions 
of exponential turnovers.  Accordingly, the assignment of synchroton photon electric field 
vectors can be made with the usual \teq{\parallel} and \teq{\perp} designation 
and textbook probabilities.  Thus
\begin{equation}
   \Pi \; \equiv\; \dover{I_{\perp}- I_{\parallel} }{I_{\perp} + I_{\parallel}}
   \; \to\; \dover{p+1}{p+7/3} \;\to\; \dover{3}{4}
 \label{eq:pol_deg}
\end{equation}
is the degree of polarization for a uniform, unturbulent field (\teq{\sigma \to 0}).

These polarization assignments are then converted into contributions to the Stokes parameters as an 
electron continues its trajectory.  If the projection of the magnetic field onto the 
{\it sky plane} \teq{(x,\, y)} lies parallel to the mean field vector \teq{\mathbf{B}_0} in the 
\teq{x}-direction, then one is at liberty to assign
\teq{Q=I, U=0} for the \teq{\parallel} state, and \teq{Q=-I, U=0} for the 
\teq{\perp} state.  Since this field orientation does not point to an observer, 
the zero circular polarization sets \teq{V=0}.  Adding components to {\bf B} in the 
\teq{z}-direction perpendicular to the sky plane does not alter this polarization, 
but it does change the synchrotron intensity, since the field vector has 
been modified.  Adding a field component in the \teq{y}-direction amounts to 
rotating the sky plane-projected field about the \teq{z}-axis through some angle 
\teq{\chi} (the {\bf position angle}), and this then mixes the Stokes parameters 
(tensor elements) so that 
\teq{Q=I \cos 2 \chi, U= I \sin 2 \chi} for a \teq{\parallel} state contribution, 
and \teq{Q=-I \cos 2 \chi, U= I \sin 2 \chi} for a \teq{\perp} state one.  These are then added 
for each step along the electron path, sampling different local field vectors and 
varying pitch angles.  As long as the local field components are inferior to the total 
sky plane-projected {\bf B}, the circularity is simply \teq{V=0}.  
The result is an integrated ensemble of Stokes \teq{Q}, \teq{U} and \teq{I} information.

For the purposes of illustration, the emission was discretized using the cubic slab structure
in a sequence of slices abutting each other along the shock face, 
but also in bins of equal thickness in the shock normal (\teq{x}) direction.  Thus, a slice 
constitutes a volume \teq{-r \leq x\leq +r, -nr/N \leq y \leq + nr/N, -r \leq z\leq +r}, and
each bin defines a rectangular prismatic rod of volume 
\teq{-nr/N \leq x\leq +nr/N, -nr/N \leq y \leq + nr/N, -r \leq z\leq +r}
(for integer \teq{n\leq N}) that is aligned 
normal to the sky plane.  Here \teq{r} is the slab half-thickness, which scales with the thickness 
\teq{R_{\rm shell}} of the SNR shell; \teq{r=10} is chosen here.  Also, \teq{N=4} is chosen 
for the purposes of illustration.  This approximates the sky plane {\it pixelation density} achievable with 
IXPE for typical remnants as afforded by its angular resolution of \teq{\sim 30''} --
see various figures in \cite[Bykov et al. (2009)]{Bykov09} for a range of intensity imagery realizable with 
X-ray polarimetry observations of turbulent SNR shells.

\begin{figure}[ht]
\begin{center}
    \centerline{\includegraphics[width=2.6in]{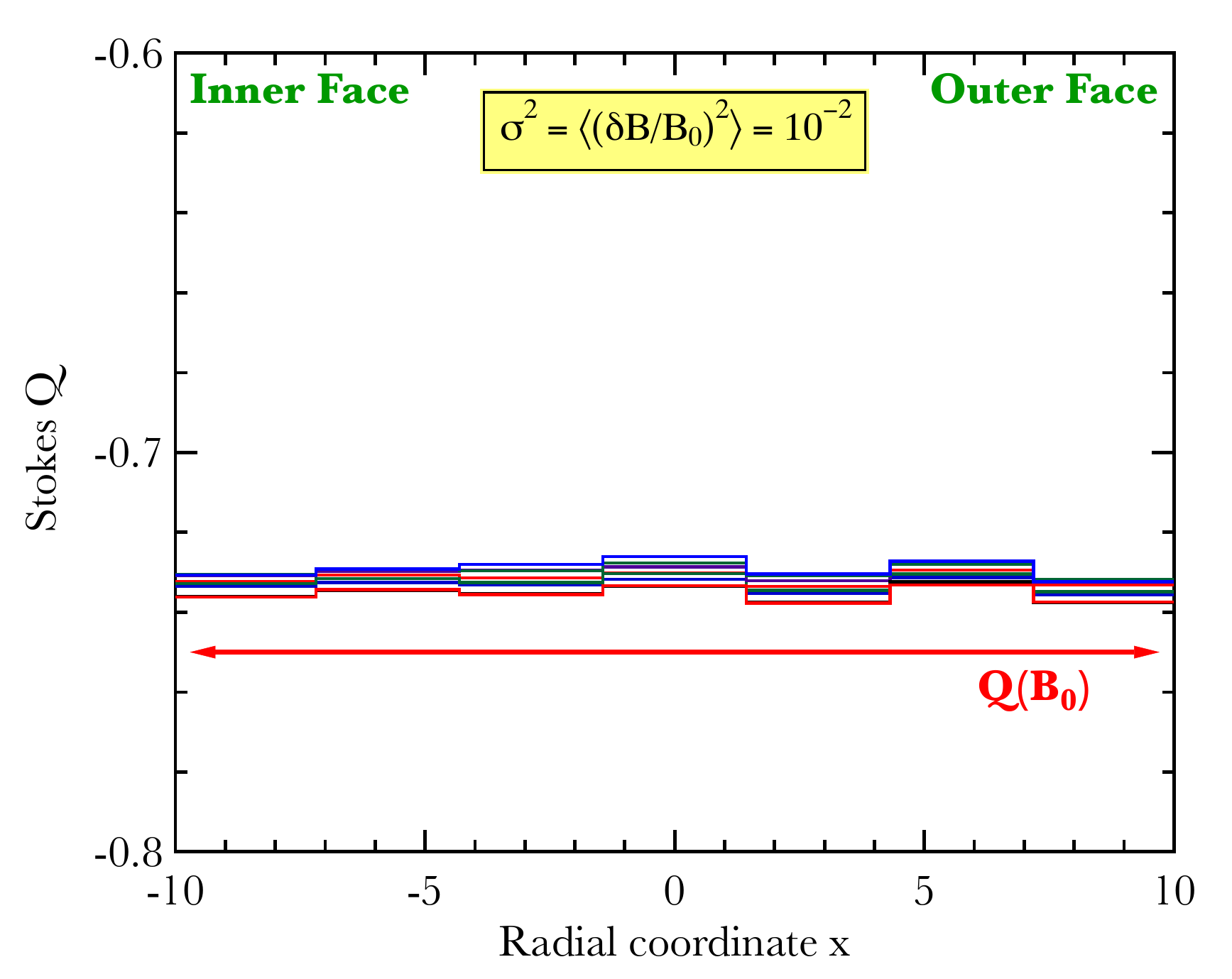} 
    \hspace{-8pt}
    \includegraphics[width=2.6in]{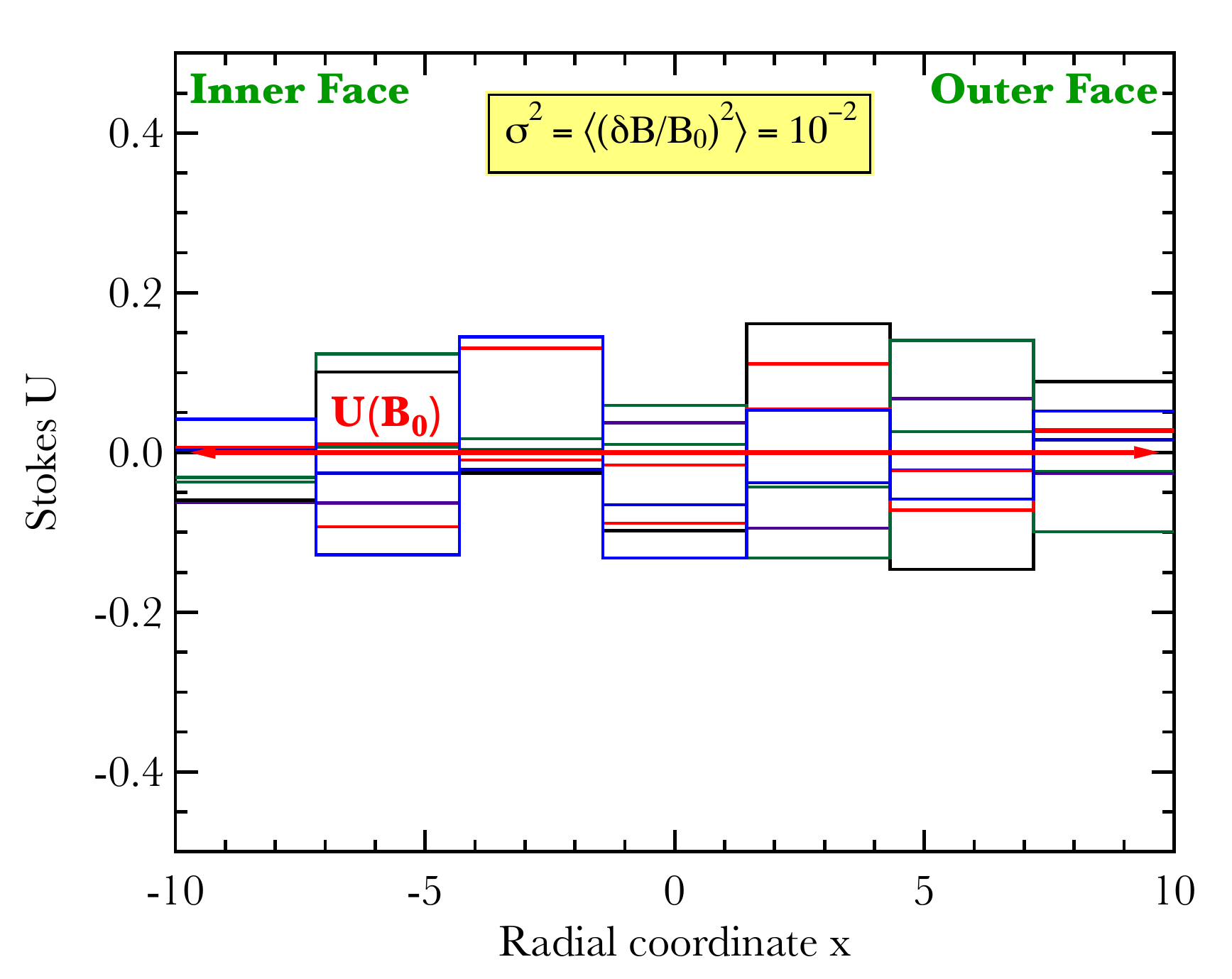}}
    \vspace*{-0.1 cm}
     \centerline{\includegraphics[width=2.6in]{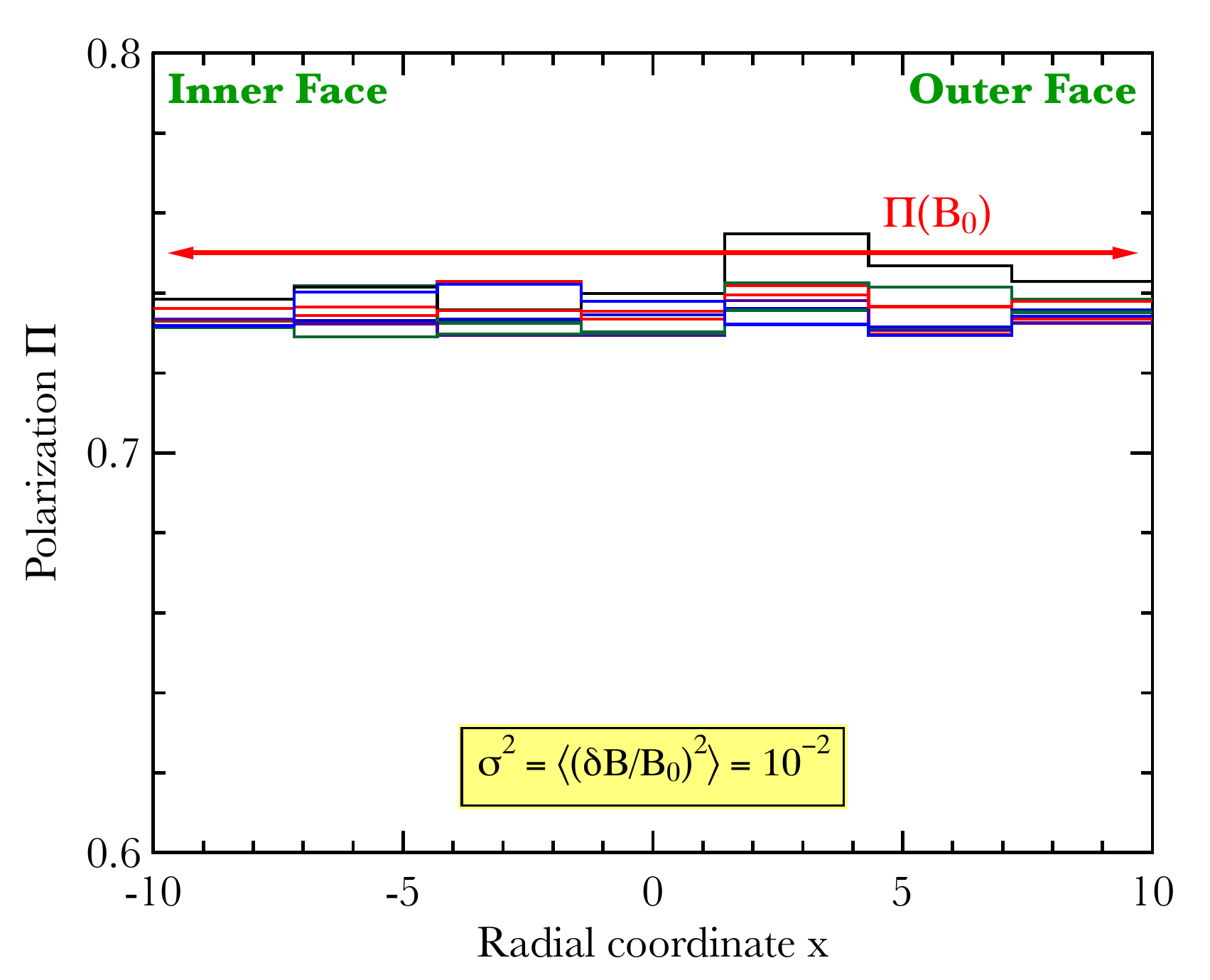} 
    \hspace{-8pt}
    \includegraphics[width=2.6in]{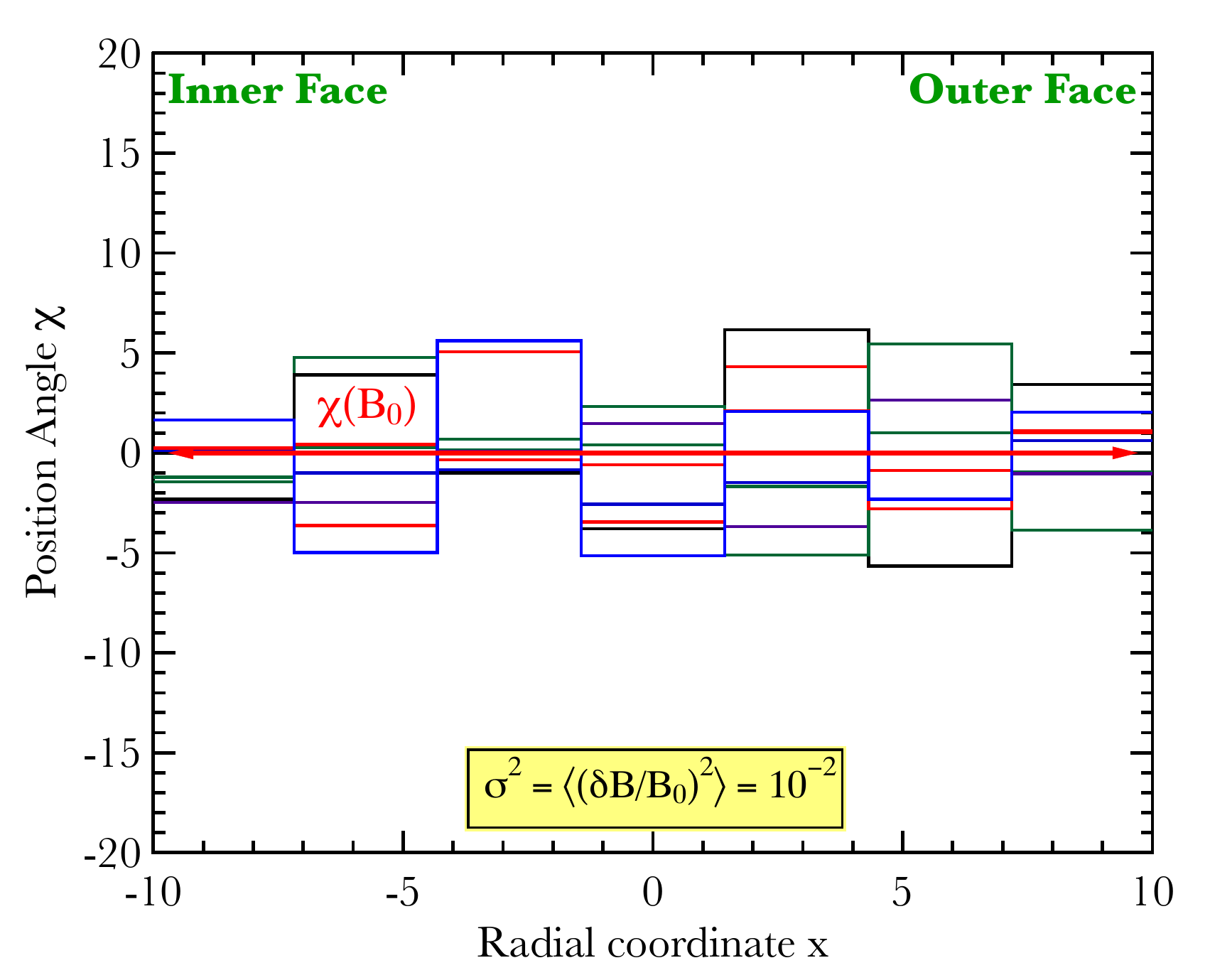}}
\caption{Synchrotron X-ray polarization quantities from power-law electrons 
\teq{n_e(\gamma) \propto \gamma^{-p}} with distribution index \teq{p=3},
for a sequence of slices of the shell that are
oriented in the \teq{x}-direction normal to the planar shock, and whose centers 
are laterally distributed with equidistant separations along the shock.  Each 
section of a histogram thus represents the polarization measure associated 
with a rectangular prism segment of a slice, with a square cross section 
in the plane of Fig.~\ref{fig1:snr_turb}. The color coding is to aid distinguishability of 
results for different slices, with color correspondence through all panels. 
The field turbulence had a Kolmogorov spectrum above the stirring 
scale \teq{\Delta x = 20/3}, at which the variance was \teq{\sigma^2 = 10^{-2}}.  
The Stokes Q and U values are depicted in the upper panels, with circularity \teq{V} being zero.  
The polarization degree \teq{\Pi} and position angle \teq{\chi = 1/2\, \arctan (U/Q)} 
are in the bottom two panels.  The values of all four quantities for a uniform field 
\teq{\mathbf{B}_0} are indicated by the red double-headed long arrows.   }
  \label{fig2:synchpol_medsig}
\end{center}
\end{figure}

\begin{figure}[ht]
\begin{center}
    \centerline{\includegraphics[width=2.6in]{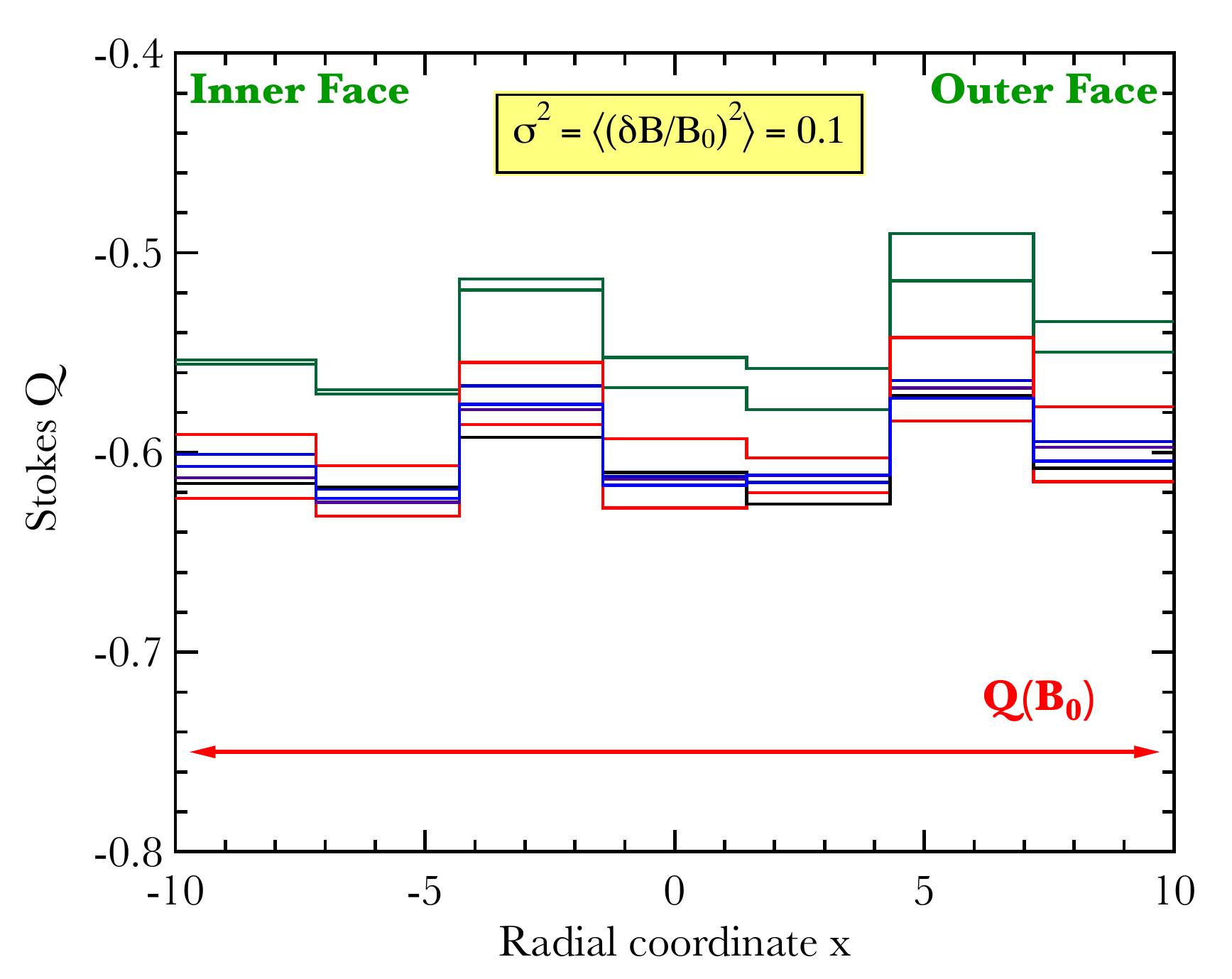} 
    \hspace{-8pt}
    \includegraphics[width=2.6in]{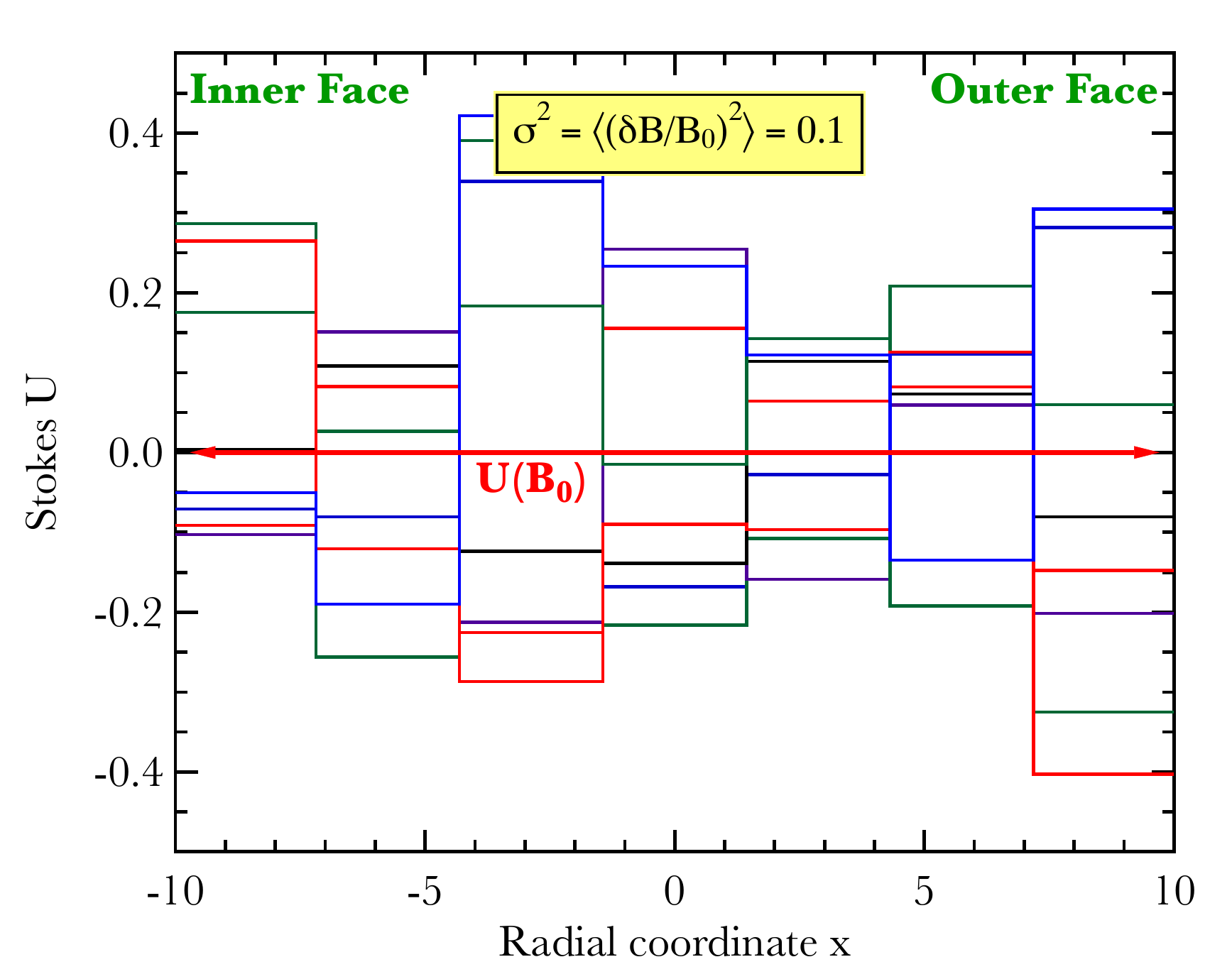}}
    \vspace*{-0.1 cm}
     \centerline{\includegraphics[width=2.6in]{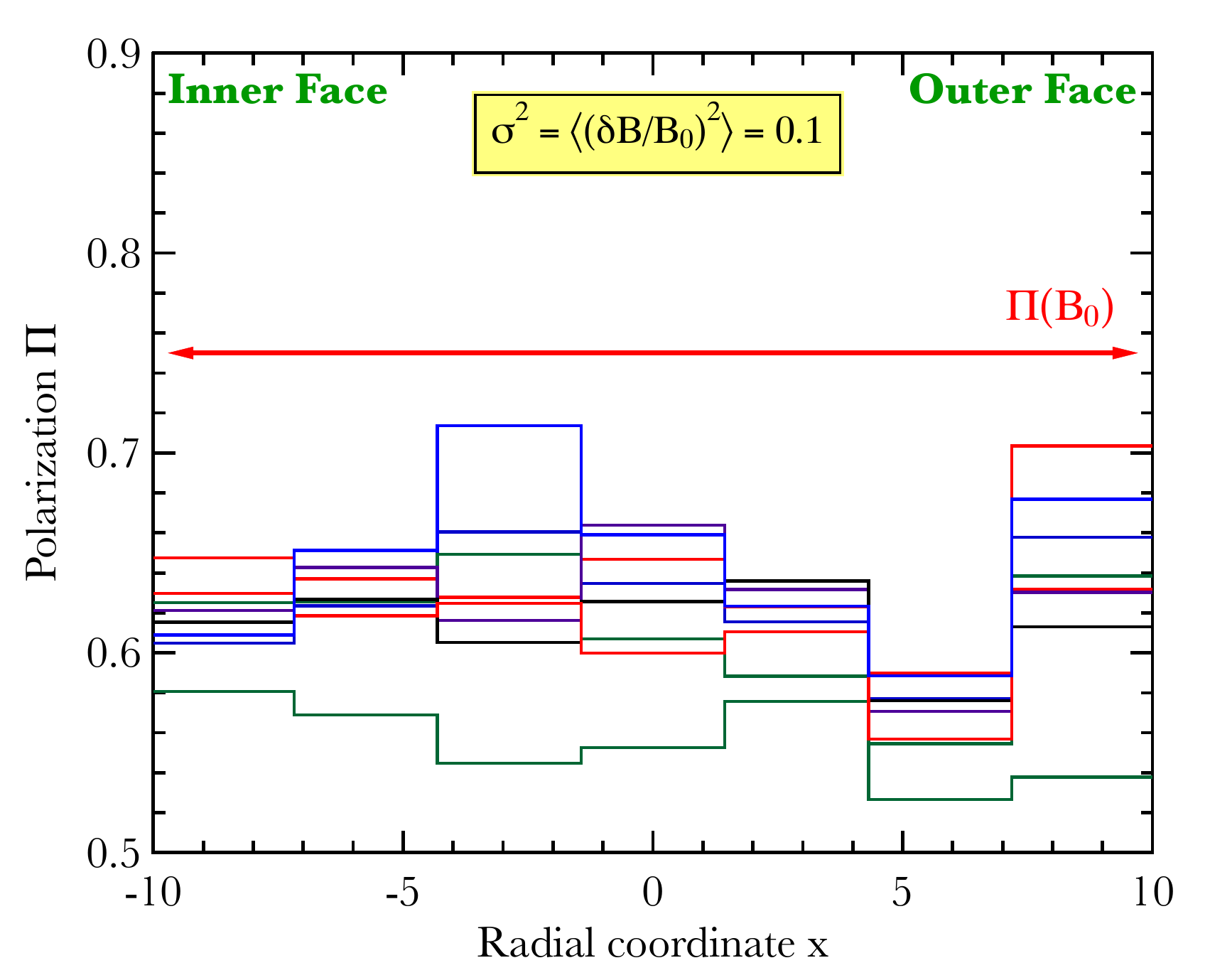} 
    \hspace{-8pt}
    \includegraphics[width=2.6in]{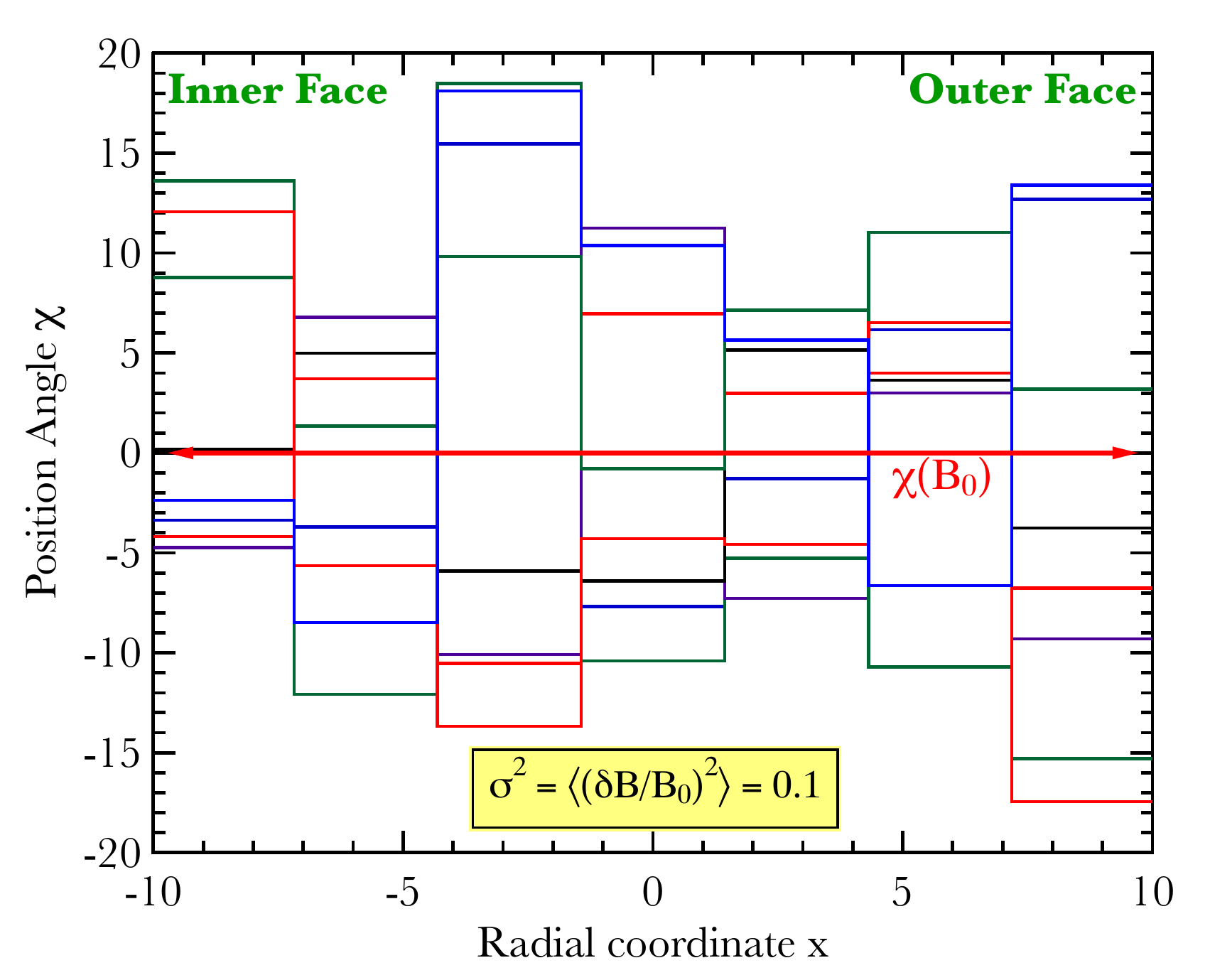}}
\caption{Synchrotron X-ray polarization quantities from power-law electrons 
as in Fig.~\ref{fig2:synchpol_medsig}, but here for a Kolmogorov spectrum with a 
higher variance of \teq{\sigma^2 = 10^{-1}} at the stirring scale.  
The enhanced depolarization and greater standard deviation in the position 
angle are evident.}
  \label{fig3:synchpol_highsig}
\vspace*{-7pt}
\end{center}
\end{figure}

The synchrotron Stokes parameter contributions from \teq{10^4} electrons injected with a 
narrow range of Lorentz factors into slab at the \teq{x=-10} boundary were recorded, added, 
and assigned to each \teq{(x,y)} pixel according to the position of each charge as it moved along its 
trajectory.  Results for two turbulence cases are ilustrated in Figs.~\ref{fig2:synchpol_medsig}
(for \teq{\sigma^2 = 10^{-2}}) and~\ref{fig3:synchpol_highsig} (for \teq{\sigma^2 = 0.1}).
The four panels each plot histograms for each slice, color-coded to enhance the 
visual distinguishability, given the overlapping/entanglement of the histograms.  The 
intermingling of such traces is expected because of the diffusive meandering of electrons 
relative to pure gyrohelixes in uniform magnetic fields.  The 
benchmark values for a uniform background field, \teq{\mathbf{B}_0} are indicated with 
the horizontal red lines with double arrowheads.  For the chosen index, the 
polarization degree \teq{\Pi = \sqrt{Q^2+U^2} /I} realizes a value of \teq{3/4}, 
and the position angle is \teq{\chi =0} for the Stokes parameter convention adopted here.
The introduction of turbulence reduces the polarization degree value as expected, with
\teq{\Pi (\sigma ) \simeq 3/4 - 3\sigma^2/2}.  The contributions of \teq{Q} 
and \teq{U} to the deviation from the uniform field value are of the same magnitude, 
but not identical.  While \teq{Q} fluctuates a little in departing from the baseline 
\teq{3/4} value, the statistical scatter in values of \teq{U} is an unambiguous 
marker of the turbulence.  The average value of \teq{U} is close to zero, but 
the standard deviation scales as \teq{\sigma^2}.  Accordingly, collecting information 
about the mean and standard deviation of the position angle \teq{\chi} affords a
second direct diagnostic on the strength of the turbulence, namely 
\teq{\vert \chi\vert \sim \sigma^2}.  This latter result is anticipated: the 
variance in polarization direction should directly scale with the variance in 
magnetic field directions.  These results obviously are obtained in the domain 
when the pixel scale is of the order of the turbulence stirring scale.  
When the stirring scale is much smaller than the polarimeter resolution scale, 
the laminar structure of the field will emerge in the results, corresponding 
to \teq{\sigma^2\ll 1} domains.

These illustrations provide key insights into the coupling between field 
turbulence and X-ray synchrotron polarization, even though the slab geometry 
does not precisely model the spatial non-uniformity and curvature of SNR 
shell environments.  The polarization degrees represented in the two Figures 
are substantially higher than those measured in the radio for SN 1006 
in Reynolds \& Gilmore (1993) and Reynoso, Hughes \& Moffett (2013),
suggesting that the variance of field turbulence on the diffusion/gyrational scales
of GeV radio-emitting electrons (much smaller than \teq{R_{\rm shell}})
must be on the order of \teq{\sigma^2\lesssim 1}.  Yet, the position angle
imagery for the VLA data on SN 1006 indicates a coherent component to the field.
The X-ray band samples much larger diffusion scales, on the order of the shell thickness.
On these lengthscales, the cosmic-ray driven magnetic field amplification models predict 
\teq{\delta B/B \sim 1}, for which the results here suggest not only 
polarization degrees \teq{\Pi < 0.4}, but that the position angle information 
in the 2--8 keV window should be highly disordered, with the variance in 
\teq{\chi} providing a direct measure of the variance in the field fluctuations.
Moreover, the coupling constant should be dependent on the binning 
or pixelation size of the data, providing constraints on the stirring scale 
for the turbulence.  We note that if IXPE measures polarization levels
in excess of 30--40\%, theorists will have to revise the 
paradigm of turbulence generation in SNR shocks.

These brief insights elicit an exciting indication of the prospects for 
advances in our understanding of MHD turbulence in supernova
remnant shells in the coming era of imaging X-ray polarimetry.

\vspace{-10pt}

\end{document}